\documentclass[%
twocolumn,
superscriptaddress,
amsmath,amssymb,
aps,
pra,
showkeys,
]{revtex4-2}

\usepackage[dvipsnames]{xcolor}

\newcommand{\ket}[1]{| #1 \rangle}
\newcommand{\bra}[1]{\langle #1 |}

\newcommand{\braket}[2]{\langle #1 | #2 \rangle}

\newcommand{\ha}{\hat{a}}
\newcommand{\had}{\hat{a}^\dagger}

\newcommand{\hH}{\hat{H}}

\newcommand{\hV}{\hat{V}}
\newcommand{\hT}{\hat{T}}

\newcommand{\hN}{\hat{N}}
\newcommand{\hn}{\hat{n}}

\newcommand{\hU}{\hat{U}}

\newcommand{\hto}{H$_2$O}
\newcommand{\Hp}{\hH_{\text{prob}}}
\newcommand{\hX}{\hat{X}}
\newcommand{\hY}{\hat{Y}}
\newcommand{\hZ}{\hat{Z}}

\newcommand{\hc}{\hat{c}}
\newcommand{\hcd}{\hat{c}^\dagger}

\newcommand{\ie}{\textit{i.e. }}

\usepackage{graphicx}
\graphicspath{{figures/}}
\usepackage{dcolumn}
\usepackage{bm}
\usepackage[hidelinks]{hyperref}
\usepackage{cleveref}
\crefname{equation}{Eq.}{Eqs.}
\Crefname{equation}{Equation}{Equations}
\crefname{figure}{Fig.}{Figs.}
\Crefname{figure}{Figure}{Figures}
\crefname{section}{Sect.}{Sects.}
\Crefname{section}{Section}{Sections}
\crefname{table}{Table}{Tables}
\crefname{appsec}{Appendix}{Appendices}
\Crefname{paragraph}{Section}{Sections}

\usepackage{booktabs}
\usepackage{multirow}
\usepackage{mathtools}
\usepackage{soul}

\begin{document}

\title{Quantum-optimal-control-inspired ans\"atze for variational quantum algorithms}

\author{Alexandre Choquette}
\affiliation{
Institut quantique \& D\'epartement de physique, Universit\'e de Sherbrooke, Sherbrooke J1K 2R1 QC, Canada
}%
\affiliation{
IBM Quantum, Zurich Research Laboratory, S\"aumerstrasse 4, 8803 R\"uschlikon, Switzerland
}%

\author{Agustin Di Paolo}%
\affiliation{
Institut quantique \& D\'epartement de physique, Universit\'e de Sherbrooke, Sherbrooke J1K 2R1 QC, Canada
}%

\author{Panagiotis Kl. Barkoutsos}%
\affiliation{
IBM Quantum, Zurich Research Laboratory, S\"aumerstrasse 4, 8803 R\"uschlikon, Switzerland
}%

\author{David S\'en\'echal}%
\affiliation{
Institut quantique \& D\'epartement de physique, Universit\'e de Sherbrooke, Sherbrooke J1K 2R1 QC, Canada
}%

\author{Ivano Tavernelli}%
\affiliation{
IBM Quantum, Zurich Research Laboratory, S\"aumerstrasse 4, 8803 R\"uschlikon, Switzerland
}

\author{Alexandre Blais}%
\affiliation{
Institut quantique \& D\'epartement de physique, Universit\'e de Sherbrooke, Sherbrooke J1K 2R1 QC, Canada
}%
\affiliation{Canadian Institute for Advanced Research, Toronto, ON, Canada}

\date{\today}

\begin{abstract}

A central component of variational quantum algorithms (VQA) is the state-preparation circuit, also known as ansatz or variational form.
This circuit is most commonly designed to respect the symmetries of the problem Hamiltonian and, in this way, constrain the variational search to a subspace of interest.
Here, we show that this approach is not always advantageous by introducing ans\"atze that incorporate symmetry-breaking unitaries.
This class of ans\"atze, that we call Quantum-Optimal-Control-inspired Ans\"atze (QOCA), is inspired by the theory of quantum optimal control and leads to an improved convergence of VQAs for some important problems. 
Indeed, we benchmark QOCA against popular ans\"atze applied to the Fermi-Hubbard model at half-filling and show that our variational circuits can approximate the ground state of this model with significantly higher accuracy and for larger systems. 
We also show how QOCA can be used to find the ground state of the water molecule and compare the performance of our ansatz against other common choices used for chemistry problems. 
This work constitutes a first step towards the development of a more general class of symmetry-breaking ans\"atze with applications to physics and chemistry problems.

\end{abstract}

\maketitle

The rise of noisy intermediate-scale quantum processors~\cite{preskill2018quantum,wright2019benchmarking} requires us to find novel algorithms designed to attenuate the effects of noise.
Variational quantum algorithms (VQA) are an example of such methods~\cite{peruzzo2014variational,mcclean2016theory}. 
These algorithms make use of a (noisy) quantum computer and a classical co-processor to minimize
a cost function specified by a problem Hamiltonian~$\Hp$. 
This minimization is achieved by preparing a state that approximates the ground state of~$\Hp$ on the quantum computer using an iterative procedure driven by the classical co-processor. 
Importantly, and thanks to the variational nature of these algorithms, this approach has been shown to potentially be resilient against noise, and well-suited to several applications including finance~\cite{zoufal2019quantum}, pure mathematics~\cite{bravo2019variational}, machine learning~\cite{schuld2019quantum,havlivcek2019supervised}, optimization problems~\cite{farhi2014quantum,barkoutsos2020improving}, quantum chemistry and materials~\cite{reiher2017elucidating,wecker2015solving,olson2017quantum,cao2019quantum,robert2019resource}, as well as quantum optics~\cite{di2019variational}.

In VQAs, the state preparation requires the parameterization of a quantum circuit, referred to as the \textit{ansatz} or \textit{variational form}, that may or may not be structured around the problem.
Recently, a considerable amount of effort has been invested in designing ans\"atze that preserve the symmetries of the problem Hamiltonian~\cite{gard2020efficient,barkoutsos2018quantum,sagastizabal2019error,ganzhorn2019gate,cade2019strategies,montanaro2020compressed}. 
The goal of symmetry-preserving strategies is to constrain the variational search to a small vector space of interest, which in principle improves the probability of convergence to the target state with fewer optimizer iterations.

In this work, we highlight shortcomings of this approach.
We then provide an ansatz that goes beyond symmetry-preserving methods by introducing a set of unitaries that break the symmetries of the problem Hamiltonian. 
To achieve this, we borrow ideas from the theory of quantum optimal control, where fast and high-fidelity operations are achieved through the addition of time-dependent symmetry-breaking terms to the Hamiltonian. 
Focusing on fermionic systems, we incorporate such terms in a time-evolution-like ansatz~\cite{wecker2015progress} to obtain the Quantum-Optimal-Control-inspired Ansatz (QOCA). 
We benchmark this approach against common ans\"atze found in the literature for the Fermi-Hubbard model and apply these ideas to the water molecule with minimal modifications. 
We find that in most cases, this method produces approximations of the target ground state that are orders of magnitude more accurate. 
To understand this improvement, we show evidence that QOCA allows for an exploration in a slightly larger Hilbert space. 

The paper is organized as follows: in~\cref{sec: ansatze}, we begin by presenting known approaches to the construction of the ansatz and then formally introduce QOCA and quantum optimal control theory in~\cref{sec:QOCA}.
We also elaborate on our strategy for the selection of symmetry-breaking terms in~\cref{sec:QOCA} and explain how these can be incorporated into a variational ansatz for the Fermi-Hubbard model in~\cref{sec: QOCA for FHM}.
Finally, we compare results obtained with the different approaches in~\cref{sec: results}.

\section{Variational ans\"atze}
\label{sec: ansatze}

In the VQA framework, a quantum processor stores a quantum state~$\ket{\psi(\bm\theta)}$ parametrized by a collection of classical variational parameters~$\bm\theta$. 
This state is prepared from a known and easily prepared reference state,~$\ket{\psi_0}$, using a quantum circuit (the ansatz) $\hU(\bm\theta)$ such that $\ket{\psi(\bm\theta)}=\hU(\bm\theta) \ket{\psi_0}$. 
The value of $\bm\theta$ is iteratively adjusted by a classical co-processor with the purpose of minimizing the cost function
\begin{equation}
        E[\bm\theta] = \frac{\bra{\psi(\bm\theta)}\Hp\ket{\psi(\bm\theta)}}{\braket{\psi(\bm\theta)}{\psi(\bm\theta)}}.
\end{equation}
Numerous variational forms~$\hU(\bm\theta)$ have been explored in the literature~\cite{kandala2017hardware,wecker2015progress,dallaire2019low,romero2018strategies,kivlichan2018quantum,barkoutsos2018quantum,woitzik2020entanglement}. 
Before introducing our approach, in this section we briefly review two widely used ans\"atze highlighting their advantages and disadvantages.

\subsection{Hardware-efficient Ansatz}
\label{sec:HEA}
The Hardware-efficient Ansatz (HEA), introduced in Ref.~\cite{kandala2017hardware}, relies on gates that are native to the quantum hardware to produce circuits of high expressibility~\cite{sim2019expressibility} and low depth.
In particular, the HEA requires the application of successive blocks of parametrized single-qubit rotations followed by a generic entangling unitary~$\hU_{\text{Ent.}}$. 
An example for~$N$ qubits is
\begin{equation}\label{eq: HEA}
        \hU_{\text{HEA}} (\bm\theta)  = \prod_{d} \hU_\text{Ent.} \prod_{n=1}^{N} R_Z^{(n)}(\theta_{n,d}^Z) R_Y^{(n)}(\theta_{n,d}^Y),
\end{equation}
where~$\bm\theta = \{ \theta_{n,d}^Z, \theta_{n,d}^Y\}$ collects all the variational parameters and~$R_{\sigma_a}^{(n)}(\theta)=\exp[-i\theta \sigma_a/2]$ denotes a single-qubit rotation of angle~$\theta$ around the~$a\in \{x,y,z\}$ axis on qubit~$n$.
~$\sigma_a $ is the corresponding Pauli matrix. 
The parameter~$d$ is the number of layers, or \textit{depth}, of the ansatz. 
Here and for the rest of this paper, we use the convention~$\prod_i^N \hU_i = \hU_N \cdots \hU_1$ for operator multiplication. 

A feature of the HEA is that it is well suited to a broad exploration of the Hilbert space since it does not purposely favor a particular symmetry sector.
This ansatz has already been experimentally implemented to prepare the ground state of small molecules~\cite{kandala2017hardware}, to simulate the folding of a few amino acid polymer~\cite{robert2019resource}, and to find the solution of classical optimization problems~\cite{barkoutsos2020improving}.
However, solving small instances of important problems does not provide a proof of scalability of the method for larger systems. 
Indeed, there is evidence that sufficiently random parametrized circuits, such as the ones produced by HEA, suffer from an exponentially vanishing gradient with the number of qubits making them more difficult to converge as the system size grows~\cite{mcclean2018barren}.

\subsection{Variational Hamiltonian Ansatz}
\label{sec:VHA}
Ans\"atze that leverage the structure of the problem can avoid the aforementioned scalability issues since they do not explore the full exponentially large Hilbert space.  
\textcite{wecker2015progress} introduced the Variational Hamiltonian Ansatz (VHA), which consists of a parametrized adaptation of the quantum circuit implementing time evolution under the problem Hamiltonian via Trotterization. 
In the VHA framework, the state-preparation unitary reads
\begin{equation} \label{eq: VHA}
        \hU_{\text{VHA}} (\bm\theta)  = \prod_{d} \prod_{j} e^{i \theta_{j,d} \hH_j },
\end{equation}
where~$\bm\theta = \{ \theta_{j,d}\}$ are the variational parameters and~$\Hp = \sum_j \hH_j$ is the problem Hamiltonian expressed as the sum of non-commuting groups of terms labeled~$\hH_j$. 
The depth~$d$ is associated with each time increment of the Trotterization of the time-evolution operator. 
If grouping the terms is done efficiently, this approach can be implemented using few variational parameters, therefore simplifying the classical optimization. 
However, depending on the complexity of the problem, circuits can be considerably longer as compared to those typically used with the HEA. 

\paragraph*{Fourier-transformed VHA (FT-VHA)}
To further reduce the number of variational parameters, it is possible to take advantage of the fact that most fermionic Hamiltonians can be written as~$\Hp = \hT + \hV$, where the diagonal bases of~$\hT$ and~$\hV$ are related through the fermionic Fourier transformation (FT)~\cite{verstraete2009quantum, ferris2014fourier, jiang2018quantum}. 
With the FT-VHA variational form, the FT is used to alternate between these bases at every Trotter step. 
In the context of quantum chemistry, this is known as the split-operator method~\cite{fleck1976time,feit1982solution}.
This idea was also recently introduced by~\textcite{babbush2018low} for the variational quantum simulation of materials.
The state-preparation unitary thereby reads
\begin{equation} \label{eq: FT-VHA}
    \begin{split}
        &\hU_{\text{FT-VHA}} (\bm\tau, \bm\nu)  \\
        &\quad= \prod_{d} \text{FT}^\dagger \left( \prod_{j} e^{i \tau_{j,d} \hat{\mathcal{T}}_{j} }\right) \text{FT} \left( \prod_{j} e^{i \nu_{j,d} \hV_j }\right),
    \end{split}
\end{equation}
where~$\bm\tau = \{ \tau_{j,d}\}$ and~$\bm\nu = \{ \nu_{j,d}\}$ are the parameters associated with~$\hat{\mathcal{T}} = \sum_j \hat{\mathcal{T}}_j = \text{FT}\, \hT\, \text{FT}^\dagger$ and~$\hV$, respectively. 
Since now both~$\hat{\mathcal{T}}$ and~$\hV$ are diagonal, they only contain terms that commute and therefore the circuit decomposition of their exponentials can be achieved exactly, which was not the case of~$\hT$ in the regular VHA.
However, this comes at the cost of the long FT circuit~\cite{verstraete2009quantum,jiang2018quantum,babbush2018low}.

Because they are built from the problem Hamiltonian, both VHA and FT-VHA respect the symmetries of the problem. 
For example, if no term of~$\Hp$ allows the number of particles to change, this quantity will be conserved in the variational state $\ket{\psi(\bm\theta)}$. 
This choice restricts the variational search to a relatively smaller subspace of the Hilbert space which, 
intuitively, can increase the performance of the VQA. 
Because of this, the VHA and FT-VHA ans\"atze are likely to perform better than HEA for large system sizes. 
However, as we show in~\cref{sec: results}, incorporating too much knowledge of the problem can also be detrimental. 

Another popular approach in quantum chemistry is the UCCSD ansatz~\cite{peruzzo2014variational} which implements the exponential of a set of single- and double-excitation operators.
Although not strictly Hamiltonian-based, this method preserves the parity symmetry of fermions and conserves the number of particles.
Despite potentially providing accurate results, the UCCSD ansatz circuits can be very deep, limiting its applicability on near-term quantum devices.

\section{Quantum-optimal-control-inspired Ansatz (QOCA)} 
\label{sec:QOCA}

To address the drawbacks of the ans\"atze discussed above, we propose an ansatz that borrows ideas from the theory of quantum optimal control~\cite{dong2010quantum,james2014quantum,khaneja2005optimal,motzoi2011optimal}, and which we therefore dub the Quantum-Optimal-Control-inspired Ansatz, or QOCA. 
The main idea behind QOCA resides in the introduction of carefully chosen symmetry-breaking unitaries into the symmetry-preserving ansatz VHA.
In this section, we begin by reviewing some of the central aspects of the theory of quantum optimal control, and then show how these ideas can be incorporated in the design of variational forms.

\subsection{Quantum optimal control}
\label{sec: QOC}
Quantum optimal control (QOC) theory describes the methods to optimally steer a quantum system from an initial state to a known final state~\cite{d2007introduction}.
Such techniques have been applied to a wide variety of problems including
the quantum control of chemical reactions~\cite{rice2001interfering,shapiro2006quantum}, spins in nuclear magnetic resonance experiments~\cite{khaneja2003optimal,khaneja2005optimal} and, more recently, to superconducting qubits~\cite{motzoi2011optimal,Heeres2017}.

In this approach, the control Hamiltonian is specified by a set of time-independent drive Hamiltonians~$\{ \hH_k \}$ whose amplitudes are parametrized by the time-dependent coefficients~$\{ c_k(t) \} \in \mathbb{R}$. 
The total Hamiltonian~$\hH(t)$ is then, in general, time-dependent such that
\begin{equation}
    \label{eq: H control}
    \hH(t) = \hH_0 + \sum_k c_k(t) \hH_k,
\end{equation}
with $\hH_0$ the free, or drift, Hamiltonian of the controlled system.
Solving the Schrödinger equation of the driven system results in the unitary~$\hU(t)$, which can propagate pure states through time as~$\ket{\psi(t)}=\hU(t)\ket{\psi(0)}$.

The system described by the Hamiltonian of~\cref{eq: H control}, defined in a Hilbert space of dimension~$n$, is said to be controllable if~$\hU(t)$ can be any matrix of~$SU(n)$.
In other words, the system is controllable if for any initial state~$\ket{\psi(0)}$, there exists a set controls~$\{c_k(t)\}$ and a time~$T>0$ for which the state~$\ket{\psi(T)}$ can be any target state of the Hilbert space~\cite{d2007introduction}. 

Quantum optimal control techniques, such as the GRAPE algorithm~\cite{khaneja2005optimal}, provide a method for designing the control pulses $c_k(t)$ to achieve a desired state preparation. 
This is usually realized by seeking the set of controls and time~$T$ that optimize a cost function characterizing the state-preparation fidelity, which may include constraints such as the control time and the maximum pulse amplitudes.

In the GRAPE algorithm, time is discretized into~$N$ increments, or pixels, of duration~$\Delta t$ such that the total evolution occurs in a time~$T=N\Delta t$. 
Using this discretization, the continuous control fields~$c_k(t)$ are now parametrized by the new constant piecewise control fields~$\mathbf{u}_k = \{u_{k,j}\}$ as
\begin{equation}\label{eq: control fields}
    c_k(t) = \sum_{j=0}^{N-1} u_{k,j} \sqcap_j(t,\Delta t),
\end{equation}
where~$\sqcap_j(t,\Delta t) \equiv \Theta(t-j\Delta t) - \Theta(t-(j+1)\Delta t)$ with~$\Theta$ the Heaviside function. 
The time evolution operator for a time~$T$ therefore reads
\begin{equation}
    \label{eq: QOC Uj}
    \hU(T) = \prod_{j=0}^{N-1} \exp \left[ -i\Delta t \left( \hH_0 + \sum_k u_{k,j} \hH_k\right) \right],
\end{equation}
and optimality is achieved by iteratively tuning the values of the discrete control fields~$\{u_{k,j}\}$.
Because this time propagator incorporates drive terms~$\{\hH_k\}$, that typically do not commute with the drift Hamiltonian, $\hU(T)$ may implement unitary operations that are distinct from that generated by the drift Hamiltonian alone. 
In standard QOC problems, fast and efficient optimization of the control fields is possible because the target state (or operation) is known. 
This is, however, not the case in the context of VQA. 
Adapting these techniques to the VQA setting therefore requires to eliminate any information about the target state from the QOC cost function, therefore making the optimization less straightforward.

\subsection{The QOCA variational form}
Building on the concept of quantum optimal control, we modify the VHA by constructing
a variational form which includes a set of \textit{drive} terms~$\{\hH_k\}$ in addition to the problem Hamiltonian~$\Hp$. 
QOCA therefore mimics time evolution under the new Hamiltonian
\begin{equation}\label{eq: Ham QOCA}
    \hH_{\text{QOCA}}(t) = \Hp + \sum_k c_k(t) \hH_k,
\end{equation}
where, by design,~$[\Hp, \hH_k]\neq 0\,\,\forall\,\, k$. 
We then construct the state-preparation circuit for QOCA by parameterizing the time-evolution-like operator
\begin{equation} \label{eq: QOCA}
        \hU_{\text{QOCA}} (\bm\theta, \bm\delta)  = \prod_{d} \left( \prod_{j} e^{i \theta_{j,d} \hH_j } \prod_{k} e^{i \delta_{k,d} \hH_k } \right),
\end{equation}
where $\Hp = \sum_j \hH_j$ and~$\bm\theta = \{ \theta_{j,d}\}$ are the problem Hamiltonian parameters.
Similar to~\cref{eq: control fields},~$\bm\delta = \{ \delta_{k,d}\}$ are the discrete drive amplitudes of the control fields~$c_k(t)$ of~\cref{eq: Ham QOCA} which we use as variational parameters.
Again,~$d$ is the depth of the ansatz and is analog to the steps in the time evolution. 

A key concept of QOCA is that the problem Hamiltonian part helps constraining the variational search to the relevant symmetry sector of the Hilbert space, while the \textit{drive} part allows the ansatz to take shortcuts by temporarily exiting this sector.
This concept is schematically drawn on~\cref{fig: ansatze}b where we illustrate possible paths in the Hilbert space for the HEA, VHA and QOCA variational forms.

In principle, one has the freedom to select any drive Hamiltonians that do not commute with~$\Hp$. 
However, it is not straightforward to predict which choice will have the most positive impact on the outcome of the VQA. 
One option is to use an adaptive approach such as the one described in Refs.~\cite{grimsley2018adapt,tang2019qubit}. 
However, in the next section we show how simple considerations can help to bound the number of interesting drive operators, and suggest which of these could be more effective.

\subsection{Which drive Hamiltonians are useful for fermions?}
\label{sec: fermionic drives}
With the objective of applying QOCA to the Fermi-Hubbard model, we consider the time-dependent fermionic Hamiltonian
\begin{equation}\label{eq: Hf}
        \begin{split}
        &\hH_f (t) = \sum_{j}(\alpha_j(t) \ha_j + \alpha_{j}^{*}(t)\had_j) \\
        &\quad + \sum_{i,j} \beta_{ij}(t)(\had_i\ha_j + \had_j\ha_i)
        + \sum_{i,j} \gamma_{ij}(t)\had_i\ha_i\had_j\ha_j,
    \end{split}
\end{equation}
where~$\had_j$ and~$\ha_j$ are fermionic ladder operators of spin-orbital~$j$ respecting the anti-commutation relations~$\{\ha_i,\had_j\}=\delta_{ij}$ and~$\{\ha_i,\ha_j\}=\{\had_i,\had_j\}=0$.
Importantly, $\hH_f (t)$ is controllable in the sense that any unitary matrix can be generated by solving its Schrödinger equation~\cite{ortiz2001quantum,ortiz2002erratum}.

We note that while the first term of $\hH_f (t)$ is unphysical since it breaks the parity symmetry of fermions, the quadratic and quartic terms occur in many physical models.
This makes $\hH_f (t)$ attractive for designing driven physically inspired ans\"atze as we are guaranteed that drive terms of form~$\alpha(t) \ha + \alpha^{*}(t)\had$ will not commute with the physical problem Hamiltonian. 
Interestingly, the use of such terms has been proposed in the context of variational error suppression~\cite{mcclean2016theory} as they may allow a variational state to re-enter a particular symmetry sector to correct for the effect of symmetry-breaking errors.

In the QOCA variational form, we propose to first write $\hH_f(t)$ keeping only the quadratic and quartic terms that also appear in~$\Hp$, along with few additional symmetry-breaking drive terms.
As in~\cref{eq: QOCA}, we then parametrize the resulting time-evolution-like operator using the associated~$\alpha(t),\,\beta(t),$ and $\gamma(t)$ coefficient as parameters.
With these choices, the QOCA variational form generates circuits that are only slightly different from those generated by the problem Hamiltonian. 

We also note that the principles of this analysis can be extended to the simulation of non-fermionic Hamiltonians, provided a controllable Hamiltonian for these systems.

\section{QOCA for the Fermi-Hubbard model}
\label{sec: QOCA for FHM}

For completeness, we start this section by reviewing
the Fermi-Hubbard model and explain how we use the QOCA variational form to prepare its ground state.
We motivate our choice of initial state, and elaborate on the selection and circuit decomposition of the drive terms.
Finally, we introduce short-QOCA, a variant of QOCA that yields shorter circuits by dropping some
terms of~$\Hp$ from the Hamiltonian that generates the regular QOCA ansatz. 

\subsection{The Fermi-Hubbard model (FHM)}
The Fermi-Hubbard model is an iconic model in the study of strongly correlated materials~\cite{hubbard1963electron}.
It describes interacting spin-$\frac{1}{2}$ fermions on a lattice where each site can be occupied by up to two particles of opposite spins.
The Hamiltonian of the FHM for~$L$ lattice sites takes the form
\begin{equation}\label{eq: FHM}
        \hH_{\text{FHM}} = \underbracket{-t \sum_{\langle i,j \rangle, \sigma}\had_{i\sigma}\ha_{j\sigma}}_{\equiv\,\hT} + \underbracket{U\sum_{i=1}^{L} \hn_{i\uparrow}\hn_{i\downarrow} - \mu \sum_{i,\sigma} \hn_{i\sigma}}_{\equiv\,\hV},
\end{equation}
where~$i,j$ are the lattice-site indices, and $\sigma=\{\uparrow,\downarrow\}$ labels the spin degree of freedom. 
In the first term,~$\langle i,j \rangle$ denotes a sum over nearest-neighbor sites, and~$\hn_{i\sigma} = \had_{i\sigma}\ha_{i\sigma}$ is the occupation operator of the spin-orbital labeled~$i\sigma$.

\begin{figure*}[]
    \includegraphics[width=2\columnwidth]{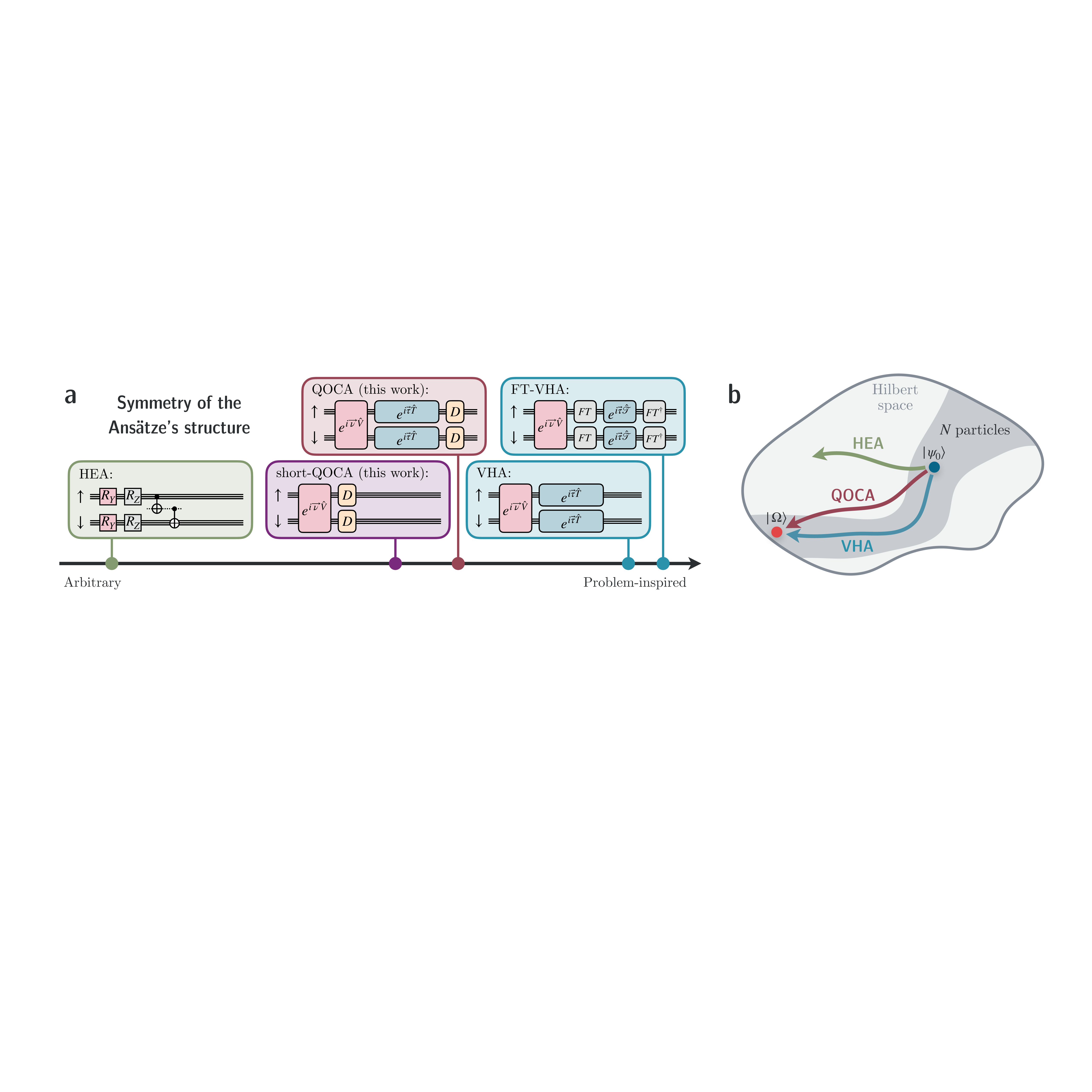}
    \caption{\label{fig: ansatze} 
    \textsf{\textbf{a}} Single circuit layer of the ans\"atze studied in this work arranged by the symmetry of their structure.
    A high symmetry means that the ansatz is completely built around the problem Hamiltonian while a low one reflects the arbitrariness of its circuit.
    We show the hardware-efficient Ansatz (HEA)~\cref{eq: HEA}, the variational Hamiltonian Ansatz (VHA)~\cref{eq: VHA}, the Fourier-transformed VHA (FT-VHA)~\cref{eq: FT-VHA}, the Quantum-Optimal-Control-inspired Ansatz (QOCA)~\cref{eq: QOCA} along with a shallower version of QOCA, the short-QOCA ansatz~\cref{eq: sQOCA}. 
    The horizontal lines represent the qubit registers that encode the spin orbitals associated with the~$\uparrow$ or~$\downarrow$ spins.
    For HEA, the entangling block is a ladder of CNOT similar to the ones in the \textit{Notation} box of~\Cref{fig: drive circuit 1}.
    For all other ans\"atze,~$\hT$ and~$\hV$ are respectively the kinetic and interaction parts of the problem Hamiltonian and~$\{\bm\tau, \bm\nu\}$ are their associated variational parameters.
    For FT-VHA, we have that~$\hat{\mathcal{T}} = \text{FT}\, \hT\, \text{FT}^\dagger$.
    The drive circuit~$D$ is defined in~\cref{eq: drive} and illustrated in~\Cref{fig: drive circuit 1}.
    \textsf{\textbf{b}} Possible paths in the Hilbert space for the HEA, VHA and QOCA variational forms.
    The initial state~$\ket{\psi_0}$ and the target state~$\ket{\Omega}$ are in the same symmetry sector containing~$N$ particles.
    Since HEA does not conserve the symmetries of~$\hH$, its path easily escapes from the fixed particle number subspace, while VHA is restricted to it.
    By introducing symmetry-breaking terms, QOCA has the ability to escape slightly from the~$N$ particles subspace to find shortcuts in Hilbert space.}
    
\end{figure*}

The first term of~\cref{eq: FHM} represents hopping between neighboring sites with amplitude~$-t$ and will generally be referred to as~$\hT$. 
This term is diagonal in momentum space if periodic boundary conditions are used, and its ground state consists of delocalized plane waves.
The second term is a non-linear, on-site Coulomb repulsion of strength~$U$, while the last term is the chemical potential. These last two terms are diagonal in the position basis and, taken together, are denoted~$\hV$.
The ground state of $\hV$ is described by wave functions localized on the sites.

A particularly interesting instance of the FHM is the half-filling regime (which occurs for~$\mu = U/2$) at intermediate coupling,~$U/t\sim 4$. 
In this regime, both~$\hT$ and~$\hV$ contribute significantly to the system's energy, thus creating competition between the localized and delocalized states of the electrons, leading to rich physics such as the Mott transition.
Because it becomes impossible to accurately treat either part of the Hamiltonian perturbatively, numerical exact diagonalization of the FHM is difficult beyond 24 lattice sites at half-filling~\cite{yamada200516}.  
As we seek to benchmark the usefulness of our variational form for all cases, we work in this particularly challenging regime. 

Despite its apparent simplicity, this model has been used to study systems ranging from heavy fermions~\cite{masuda2015variational} to high-temperature superconductors~\cite{guillot2007competition,kaczmarczyk2013superconductivity}. 
As a result, it is an interesting problem to benchmark near-term quantum computers~\cite{dallaire2020application}, and a useful performance test for variational ans\"atze. 
For these reasons, variational quantum algorithms have already been used to find the ground state of the  FHM, for example using the HEA variational form~\cite{wilson2019optimizing}, the VHA~\cite{wecker2015progress,reiner2019finding,verdon2019learning,cade2019strategies,montanaro2020compressed}, and other symmetry-preserving ans\"atze~\cite{dallaire2019low,sokolov2020quantum,cade2019strategies,xu2020test,dallaire2020application}.

\subsection{Encoding and parametrization of the ans\"atze}
We use the Jordan-Wigner (JW) transformation to encode fermionic Fock states into qubits registers, as detailed in~\Cref{sec: JW}.
Moreover, we work in real space and order the basis vectors for the~$2L$ spin orbitals as~$\ket{f_{1\uparrow}\dots f_{L\uparrow} ; f_{1\downarrow}\dots f_{L\downarrow}}$ with~$f_p\in \{0,1\}$ the occupation of orbital~$p$. 

Using this purely conventional choice, in~\cref{fig: ansatze}a we schematically draw one layer of the circuits implementing the different ans\"atze discussed above and arranged by the symmetry of their structure.
A highly symmetric ansatz is completely built around~$\Hp$ while a weakly symmetric construction is arbitrary with respect to the problem. 

To parametrize these circuits, we consider two possible strategies: one corresponding to full parametrization of the single- and two-qubit gates and the other having a number of parameters that only grows with the depth of the ansatz, but not with the number of qubits.
Whenever used, the latter is specified with the label `scalable'.
Both strategies are elaborated on in~\Cref{sec: parametrization} and details of the numerical simulation are presented in~\Cref{sec: numerical}.

\subsection{Initial state}
In general, the performance of VQAs strongly depends on the choice of initial state and variational parameters. 
The initial state acts as an educated guess to the target state and is often chosen such as to be easily computable classically. Moreover, because the initialization stage of a variational algorithm should be straightforward or otherwise be treated as a separate routine~\cite{hadfield2019quantum}, we are interested in benchmarking the performance of the QOCA variational form for the simple initial state
\begin{equation}
    \ket{\psi_0} = H^{\otimes N}\ket{0} = \ket{+}^{\otimes N},
    \label{eq: initial state}
\end{equation}
where~$H$ is the Hadamard gate.
In addition to being easy to prepare, this initial state corresponds to half-filling and zero total spin, placing it in the same symmetry sector as the target state.

While this choice allows us to demonstrate the usefulness of the QOCA variational form given unstructured, simple initial conditions, we also show how the convergence can be improved further by using the ground state of the non-interacting FHM fixing~$U=\mu=0$ in~\cref{eq: FHM} as initial state. 
More details on how to prepare this more complex state are provided in~\Cref{app: initial states}.

\subsection{Drive Hamiltonians}
\begin{figure}[]
    \includegraphics[width=1\columnwidth]{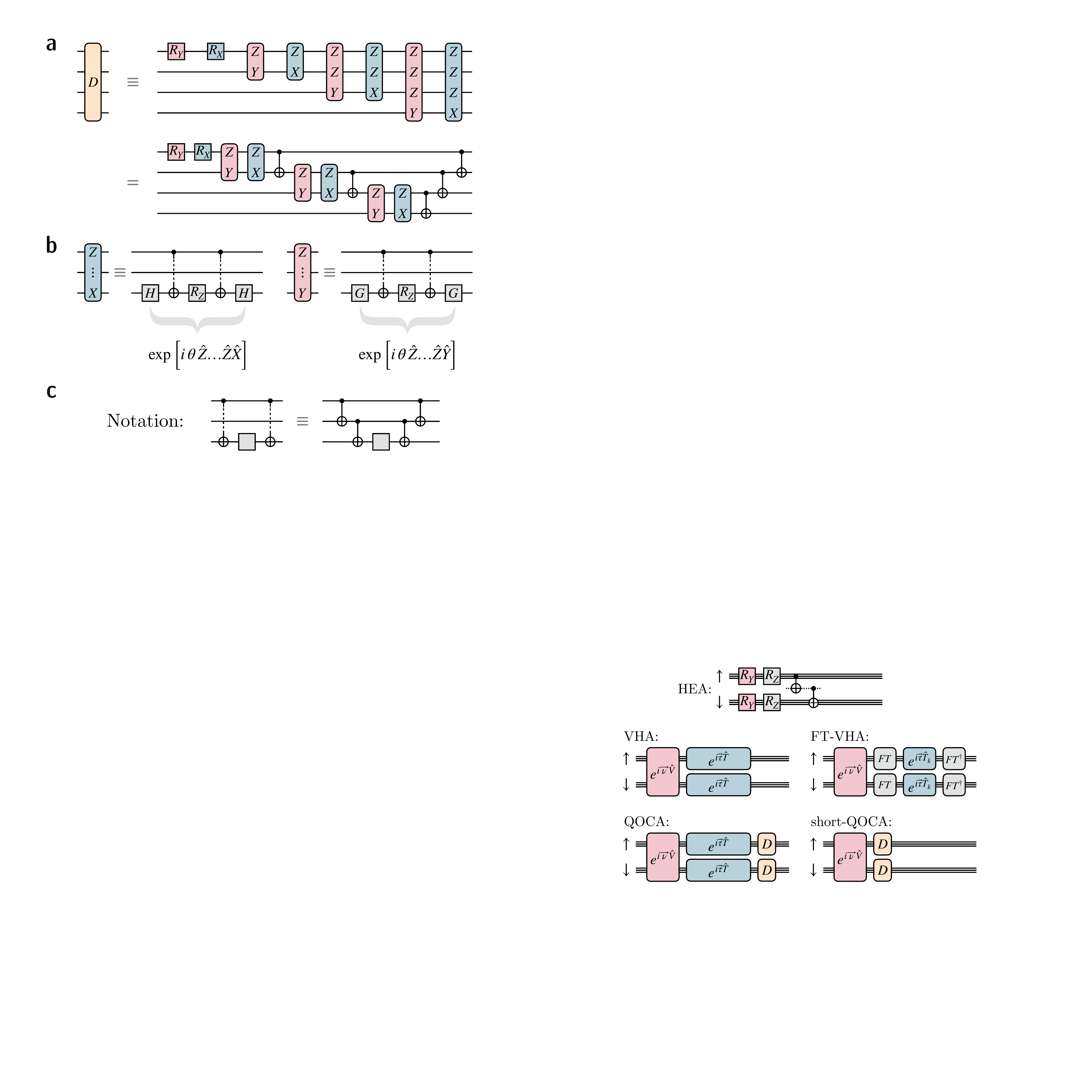}
    \caption{\label{fig: drive circuit 1} 
    \textbf{a} Circuit decomposition of the drive~\cref{eq: drive} used for QOCA. 
    This circuit generalizes to any number of qubits by appending more~$\hZ\dots\hZ\hY$ and~$\hZ\dots\hZ\hX$ multi-qubit gates at the end.
    We also show the circuit compiled to one- and two-qubit (CNOT) gates.
    \textbf{b} shows a decomposition of the multi-qubit gates based on a conventional approach to decompose exponentials of Pauli strings into circuits of CNOTs described in~\cite{whitfield2011simulation}.
    The transformation~$H=(\hX+\hZ)/\sqrt{2}$ is the Hadamard gate which changes between the~$\hX$ and~$\hZ$ bases and~$G=(\hY+\hZ)/\sqrt{2}$ is the equivalent transformation between the~$\hY$ and~$\hZ$ bases. 
    The angles of the rotations~$R_a(\theta)=\exp [-i\theta \hat\sigma_a /2]$ are the variational parameters, where~$\hat\sigma_a$ is a Pauli matrix. 
    \hbox{\textbf{c} shows} the notation shortcut used for the ladders of CNOTs.}
\end{figure}

With the goal of reducing the number of variational parameters, we fix $\alpha_j(t)$ to~$1$ and~$i$ in~\cref{eq: Hf} leading to 
\begin{align}
    \label{eq: driveX}
    \hH_1 &= \sum_{j=1}^{L} (\had_j+\ha_j),\\
    \label{eq: driveY}
    \hH_2 &= \sum_{j=1}^{L} i (\had_j-\ha_j).
\end{align}
We moreover obtain the drive equations for a spinless system and independently apply the resulting circuit to the two subspaces corresponding to the spin projections up and down for all sites.
Performing the JW transformation on~\cref{eq: driveX,eq: driveY}, we find
\begin{align}
    \label{eq: driveX JW}
    \hH_1 &\mapsto \sum_{j=1}^{L}  \hX_j\, \bigotimes_{l<j}\,\hZ_l,\\
    \label{eq: driveY JW}
    \hH_2 &\mapsto \sum_{j=1}^{L}  \hY_j\, \bigotimes_{l<j}\,\hZ_l,
\end{align}
where~$\hX,\,\hY$ and $\hZ$ are Pauli matrices.
To incorporate these expressions into the QOCA variational form~\cref{eq: QOCA}, we perform a first-order Trotter-Suzuki decomposition, arriving at the circuit equation for the~$d$th layer of the ansatz,
\begin{equation}\label{eq: drive}
    \begin{split}
        &\prod_{k=1,2} e^{i \delta_{k,d} \hH_k } \\
        &\quad\approx \prod_{j=1}^L \exp \left[ i \delta_{1,d}\, \hX_j\, \bigotimes_{l<j}\,\hZ_l \right] \, \exp \left[ i \delta_{2,d}\, \hY_j\, \bigotimes_{l<j}\,\hZ_l \right] ,
    \end{split}
\end{equation}
where~$ \{\delta_{k,d}\}$ are the variational parameters associated with the~$k$th drive term of that layer. 
A schematic of the circuit implementing~\cref{eq: drive} for 4 qubits is illustrated in~\cref{fig: drive circuit 1} where we also show a compiled version in terms of CNOTs.

\subsection{The short-QOCA variational form}
\label{sec: short-QOCA}
One drawback of QOCA is that, depending on the form of the drive, the corresponding quantum circuits can be long.
Here we demonstrate a practical approach for the reduction of the circuit depth without compromising the performance.

Because the drive~$D$ in~\cref{fig: drive circuit 1} and the kinetic part of the FHM~\cref{eq: FHM} are both block-diagonal in the spin degree of freedom, we 
chose to
remove the latter term, which is also costly in terms of two-qubit gates, arriving to the simplified form of the ansatz
\begin{equation} \label{eq: sQOCA}
    \hU_{\text{sQOCA}} (\bm\nu, \bm\delta)  = \prod_{d} \left( \prod_{j} e^{i \nu_{j,d} \hV_j } \prod_{k} e^{i \delta_{k,d} \hH_k } \right),
\end{equation}
where~$\hV = \sum_j \hV_j$ is the on-site interaction part of the Fermi-Hubbard Hamiltonian~\cref{eq: FHM} and~$\bm\nu = \{\nu_{j,d}\}$ are the associated variational parameters.
We refer to this simplified version of the QOCA variational form as \textit{short-QOCA}, see~\cref{fig: ansatze}.

\section{Numerical results}
\label{sec: results}

In this section, we compare results obtained from numerical simulations of QOCA and short-QOCA for the Fermi-Hubbard model, and contrast these results with those obtained with the other ans\"atze discussed in this article. 
As an illustration of the use of QOCA beyond the Fermi-Hubbard model, we also present a comparison of the performance of this ansatz over a hardware-efficient approach and the UCCSD ansatz for a 12-qubit representation of the \hto\ molecule.

Throughout this section, we use the fidelity with respect to the target state~$\ket{\Omega}$ (i.e.~ground state of the FHM or of the water molecule) as defined by 
\begin{equation}\label{eq: Fid}
    \text{Fidelity} = |\braket{\psi (\bm\theta)}{\Omega}|^2,
\end{equation}
to quantify the quality of the variational state~$\ket{\psi (\bm\theta)}$.

\subsection{Fermi-Hubbard model}
We consider $2\times2$ (8 qubits) and $2\times3$ (12 qubits) lattices of the Fermi-Hubbard model
at half-filling
with open boundary conditions. We note that the former configuration can also be seen as a periodic~$1\times 4$ chain. 
This allows us to compare with the FT-VHA variational form, as the fermionic Fourier transform on which this approach relies is defined for periodic boundary conditions. 
Importantly, we find that for smaller systems such as the four-qubits~$2\times 1$ dimer, all ans\"atze converge in a few tens of iterations on the ground-state energy with a precision of $< 10^{-7}$ using a single ansatz layer,~$d=1$, except for the HEA which requires two layers. 

\paragraph{Comparing the ans\"atze} 

\begin{figure}[]
    \includegraphics[width=1\columnwidth]{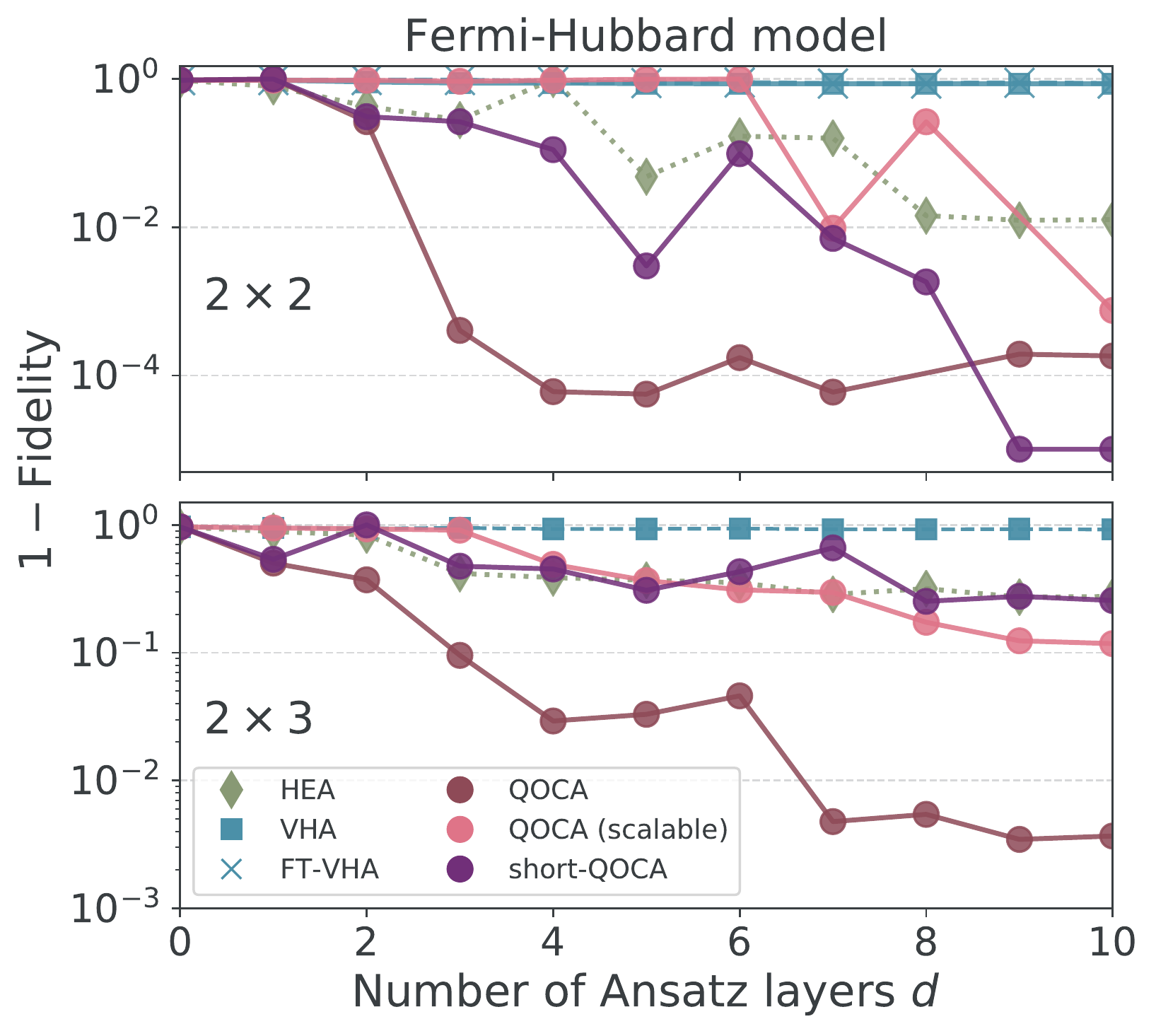}
    \caption{\label{fig: fid_vs_d} 
    Final variational state infidelities with respect to the target state as a function of the number of layers~$d$ of the variational forms of this work. 
    Top panel is for a~$2\times 2$ plaquet\-te while the bottom panel is a~$2\times 3$ system without periodic boundary conditions. 
    The initial state is~$\ket{+}^{\otimes N}$ for all cases. 
    Data at~$d=0$ corresponds to the initial state alone, which has a fidelity of~$0.035$ with the target state. 
    Unless specified, all ans\"atze are fully parametrized according to~\Cref{sec: full param}.}
\end{figure}

\begin{table}[]
    \centering
    \caption{Maximum fidelities with respect to the ground state of the FHM, attained for~$d$ ansatz layers, each requiring a number~$n_\theta/d$ of variational parameters and~$n_{\text{CX}}/d$ CNOTs per layer.
    The latter estimate assumes an all-to-all connectivity and the same compiling procedure is used for all ans\"atze.}
    \label{tab: max fids}
    \begin{tabular}{@{}lllcccc@{}}
    \toprule
    \multicolumn{2}{l}{}                                                                                                                 & Hubbard model                        & Max Fid. & $d$ & $n_\theta/d$ & $n_{\text{CX}}/d$ \\ \midrule
    \multirow{6}{*}{\rotatebox[origin=c]{90}{$2\times 2$}} & \multicolumn{1}{l|}{\multirow{6}{*}{\rotatebox[origin=c]{90}{(8 qubits)}}}  & \multicolumn{1}{l|}{HEA}             & $0.9876$ & 9   & 16        & 7                  \\
                                                           & \multicolumn{1}{l|}{}                                                       & \multicolumn{1}{l|}{VHA}             & $0.1343$ & 8   & 8         &  56                  \\
                                                           & \multicolumn{1}{l|}{}                                                       & \multicolumn{1}{l|}{FT-VHA}          & $0.1315$ & 7   & 8         &  120                 \\
                                                           & \multicolumn{1}{l|}{}                                                       & \multicolumn{1}{l|}{QOCA}            & $0.9999$ & 4   & 16         &  88                 \\
                                                           & \multicolumn{1}{l|}{}                                                       & \multicolumn{1}{l|}{QOCA (scalable)} & $0.9992$ & 10  & 5         &  88                 \\
                                                           & \multicolumn{1}{l|}{}                                                       & \multicolumn{1}{l|}{short-QOCA}      & $0.9999$ & 9   & 12        & 40                  \\ \midrule
    \multirow{5}{*}{\rotatebox[origin=c]{90}{$2\times 3$}} & \multicolumn{1}{l|}{\multirow{5}{*}{\rotatebox[origin=c]{90}{(12 qubits)}}} & \multicolumn{1}{l|}{HEA}             & $0.7276$ & 10  & 24        & 11                  \\
                                                           & \multicolumn{1}{l|}{}                                                       & \multicolumn{1}{l|}{VHA}             & $0.0804$ & 10  & 13        &  116                 \\
                                                           & \multicolumn{1}{l|}{}                                                       & \multicolumn{1}{l|}{QOCA}            & $0.9965$ & 9   & 25        & 172                  \\
                                                           & \multicolumn{1}{l|}{}                                                       & \multicolumn{1}{l|}{QOCA (scalable)} & $0.8822$ & 10  & 6         &  172                 \\
                                                           & \multicolumn{1}{l|}{}                                                       & \multicolumn{1}{l|}{short-QOCA}      & $0.7476$ & 8   & 18        &  68                 \\ \bottomrule
    \end{tabular}
\end{table}

For systems with four and six fermionic sites, we observe important variations in the ability of the different ans\"atze to converge to the ground state energy. 
This is illustrated in~\cref{fig: fid_vs_d} which shows, for all ans\"atze, the final state infidelity as a function of the number of ansatz layers,~$d$, initialized with the simple half-filled state of~\cref{eq: initial state}. 
The maximum fidelities achieved for all ans\"atze are reported in~\cref{tab: max fids} along with resource counts using a circuit compilation in terms of CNOTs.

We first note that VHA and FT-VHA perform poorly for both system sizes and that their performance does not improve with the addition of more entangling layers, \ie increasing~$d$. 
Because these ans\"atze are particle-number conserving, this observation
suggests that VHA and FT-VHA may not efficiently search over all states of fixed particle number in the variational landscape, as was originally proposed. 
Moreover, since FT-VHA performs similarly to VHA for the~$2\times 2$ system, we also conclude that alternating bases with the fermionic Fourier transform does not yield superior results for these lattice sizes. 

Interestingly, QOCA systematically reaches the ground state of the Fermi-Hubbard model with significantly more accuracy than VHA for both system sizes, indicating that the additional symmetry-breaking terms help the convergence. 
This advantage persists even when drastically reducing the number of variational parameters from 16 to 5 per layer in the case of the scalable parametrization of QOCA, which converged with~$0.9992$ fidelity at~$d=10$ for the~$2\times 2$ system.
The hardware-efficient approach also performs better than VHA, although it uses considerably more parameters than all other ans\"atze given it generally requires more layers to achieve similar performances.
It is unclear how one might reduce that number to a favorable scaling. 

Data obtained with the short-QOCA variational form show that the QOCA circuits can be substantially shortened by removing more than half of the two-qubit gates at every step without much compromise on the performance for small systems.
In fact, for the~$2\times 2$ Hubbard model, a fidelity of~$0.9999$ is achieved with~$9$ layers of this ansatz. 

With improved fidelities for shallower circuits which use fewer variational parameters than standard approaches, we find that QOCA provides significant gain with respect to other common ans\"atze.

\paragraph{The benefits of breaking symmetries}
\label{sec: breaking syms}

\begin{figure}[]
    \includegraphics[width=1\columnwidth]{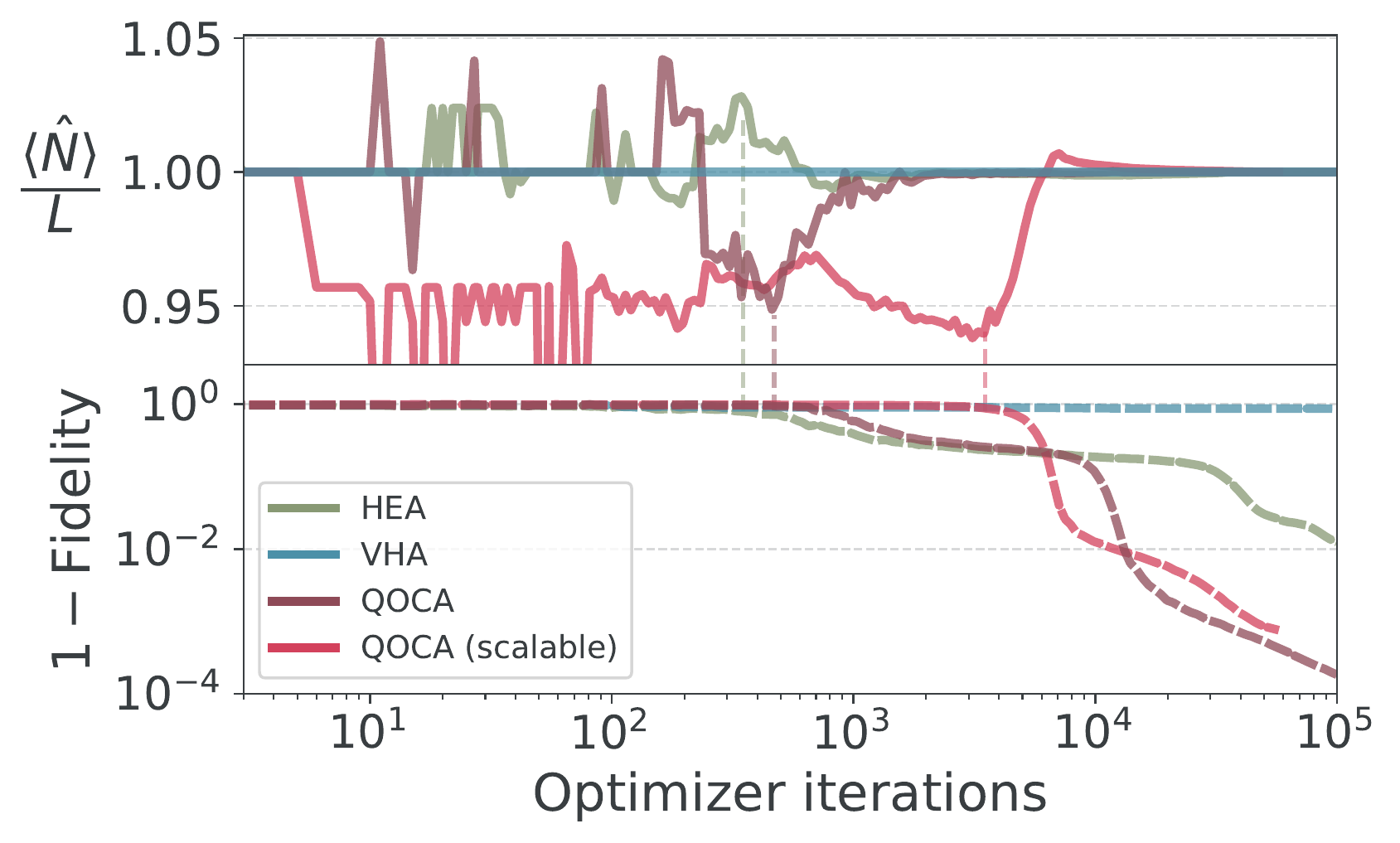}
    \caption{\label{fig: NvsIT} 
    Top: Average number of particles per lattice site in the variational state at every iteration of the VQA routine. 
    $\langle\hN\rangle = \sum_{i,\sigma} \langle \hn_{i\sigma}\rangle$ is the total occupation and~$L$ is the number of sites.
    Bottom: corresponding variational state infidelity,~$1-|\braket{\psi (\bm\theta)}{\Omega}|^2$, with respect to the ground state of the Fermi-Hubbard model~$\ket{\Omega}$. 
    The results are for a~$2\times 2$ system and the initial state is~$\ket{+}^{\otimes N}$ for all ans\"atze. 
    Runs for ansatz depth~$d=9$ was used for HEA and~$d=10$ for the others, but this behavior is observed for most~$d$.}
\end{figure}

\Cref{fig: NvsIT} shows the evolution of the average number of particles per lattice site (top panel) and the infidelity of the variational state with respect to the target state (bottom panel) throughout the optimization process for the same simulations as in~\cref{fig: fid_vs_d}.

Focusing first on the top panel we first note that, because the initial state~$\ket{+}^{\otimes N}$ is half-filled, all variational states begin in the correct particle-number symmetry sector of the Hilbert space with~$\langle\hN\rangle/L=1$. 
Because VHA does not contain terms that allow the particle number to change, this quantity is observed to be constant throughout the optimization. 
We hypothesize that the poor performance of this ansatz in reaching the ground state is caused by the inability of this variational form to overcome local minima in parameter space.

In contrast, with their particle-non-conser\-ving drive terms, both parametrizations of QOCA allow the average site occupancy to deviate from~$\langle\hN\rangle/L=1$ as the drive angles are being tuned away from zero by the optimizer. 
As seen in \cref{fig: NvsIT}, this can lead to the sharp features observed in the first few $\sim10^{2}$ iterations as the classical optimizer can initially overweight the value of individual terms. 
Over the full optimization, the number of particles deviates only slightly from the target value~$\langle\hN\rangle/L=1$ with changes of only $\sim 5\%$ of the site occupancy. 
This is an indication that the symmetry-breaking terms in QOCA allow the ansatz to explore a Hilbert space that is slightly larger than the manifold of fixed particle number. 
Nevertheless, we find that these relatively small excursions out of the target symmetry sector can significantly ease convergence of the VQA.
Indeed, we observe that the onset of the return to the target symmetry sector, as indicated by the vertical dashed lines in \cref{fig: NvsIT} is often associated with the abrupt descents in the infidelity, which may indicate that regions of steep gradients in parameter space are found. 

This behavior is also observed for the hardware-efficient ansatz of~\cref{eq: HEA} which also, does not preserve the symmetries of~$\Hp$. 
This phenomenon is not particular to the realizations displayed in the figure, and it is also observed for other system sizes and initial states.

We note, however, that these desired regions in parameter space would never be found if an error-mi\-ti\-gation technique based on symmetry verification were employed~\cite{bonet2018low,sagastizabal2019experimental}. 
Indeed, in these schemes the variational states are post-selected after the energy measurements only if they conserve desired symmetries of the target state.
However, other strategies for error mitigation remain applicable~\cite{temme2017error,endo2018practical,kandala2018extending}.

\paragraph{Initial state}

\begin{figure}[]
    \includegraphics[width=1\columnwidth]{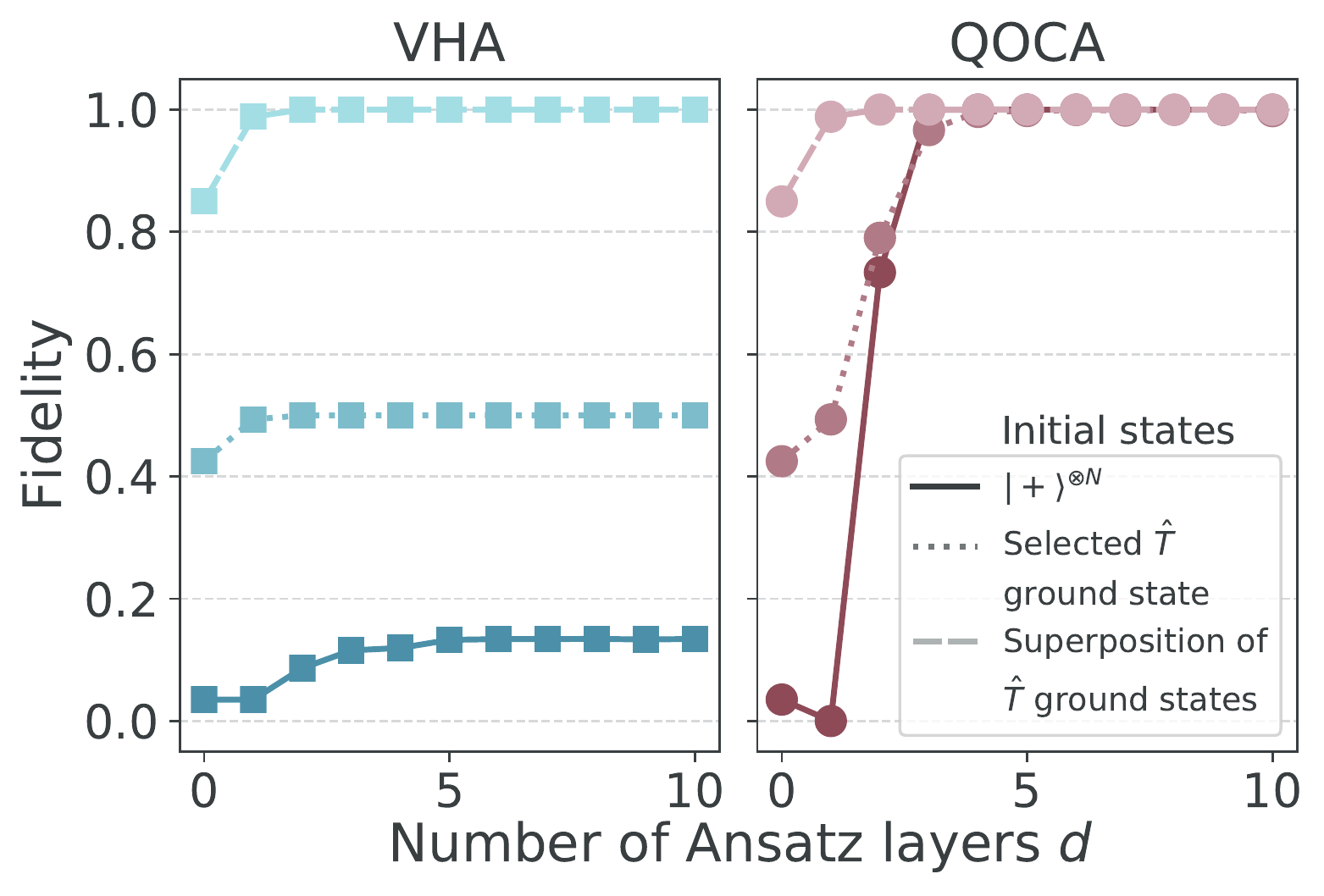}
    \caption{\label{fig: fid_vs_d_IS} 
    Variational state fidelities with respect to the ground state of the~$2 \times 2$ FHM as a function of the number of ansatz layers,~$d$, for the VHA and QOCA variational forms. 
    Results with three initial states are presented: (solid) Hadamard gates on every qubit~$\ket{+}^{\otimes N}$, (dotted) a selected ground state of~$\hT$ corresponding to~$\ket{\Omega_T^{(1)}}$ of~\Cref{app: initial states}, and (dashed) a superposition of ground states of~$\hT$ corresponding to~$\ket{\Omega_T}$ of~\Cref{app: initial states}.}
\end{figure}

Because it provides a simple setting to benchmark the performance of the different ans\"atze, we have so far considered only the single easily-prepared initial state of \cref{eq: initial state}.
Improved approximation to the ground state can, however, be obtained if a more structured initial state is considered although at the price of more complex state preparation circuits.

In~\cref{fig: fid_vs_d_IS}, we compare the performance of the VHA and QOCA variational forms on the~$2\times 2$ lattice with the following initial states of increasing complexity: i) the simple state $\ket{+}^{\otimes N}$, ii) one of the degenerate ground states of~$\hT$ labelled $\ket{\Omega_T^{(1)}}$ in~\Cref{app: initial states}, and iii) the superposition of ground states of~$\hT$ labelled $\ket{\Omega_T}$ of~\Cref{app: initial states}.

While the final variational state obtained with VHA strongly depends on the initial state, QOCA systematically achieves convergence with fidelity~$>0.9999$, regardless of the initialization choice. 
Again because of its ability to move between symmetry sectors, these results illustrate QOCA's robustness to simple, unstructured, initial conditions that can have very small overlaps with the target ground state. 
For the two variational forms, using a superposition of the degenerate ground states of~$\hT$ as initial state (dashed lines) leads to convergence with fewer entangling layers. 
This, however, comes at the cost of significantly increasing the complexity of the initialization stage of the VQA (see~\Cref{app: initial states}).

\subsection{Proof-of-principle implementation of the \hto\ molecule}

\begin{table}[]
    \centering
    \caption{\label{tab: max fids H2O} 
    Maximum fidelities obtained with~$d$ entangling layers,~$n_\theta/d$ variational parameters and~$n_{\text{CX}}/d$ CNOTs per layer and different initial states for the QOCA, HEA and UCCSD variational forms applied to the water molecule. 
    The initial states are either the Hartree-Fock (HF) approximation to the ground state or the equal superposition of all basis states~$\ket{+}^{\otimes N}$.
    Again, the gate count estimate assumes an all-to-all connectivity and the same compiling procedure is used for all ans\"atze.}
    \begin{tabular}{lccccc}
    \toprule
    Water molecule  & (12 qubits) &  &  \\ \midrule
     & Initial state\  &\ Max Fid.\ & $d$ & $n_\theta/d$ & $n_{\text{CX}}/d$ \\ \hline
    \multicolumn{1}{l|}{QOCA} & $\ket{+}^{\otimes N}$ & $0.9742$ & 1 & 23 & 108 \\
    \multicolumn{1}{l|}{} & $\ket{+}^{\otimes N}$ & $0.9931$ & 5 & 23 & 108 \\
    \multicolumn{1}{l|}{} & HF & $0.9735$ & 1 & 23 & 108 \\
    \multicolumn{1}{l|}{} & HF & $0.9917$ & 7 & 23 & 108 \\
    \multicolumn{1}{l|}{HEA} & $\ket{+}^{\otimes N}$ & $0.9820$ & 8 & 24 & 11 \\
    \multicolumn{1}{l|}{UCCSD} & HF & $0.9748$ & 1 & 8 & 528 \\
    \bottomrule
    \end{tabular}
\end{table}

The previous section illustrates how QOCA can approximate the ground state of the FHM with systematically more accuracy than other ans\"atze even when faced with unstructured initial conditions.
In order to investigate the broader applicability of this method, we now benchmark the QOCA variational form on a quantum chemistry problem.
As a proof-of-principle test, we consider the H$_2$O molecule in its equilibrium configuration.
Because we freeze the core orbitals, this problem maps to 12 qubits using the STO3G basis set.
The Hamiltonian is obtained using the PySCF driver as provided in Qiskit Chemistry~\cite{Qiskit}. 
We compare the performance of QOCA against HEA together with the well-known chemistry-inspired UCCSD ansatz~\cite{peruzzo2014variational}. 
Because the Hamiltonian describing the water molecule has significantly more terms than the FHM, directly implementing Hamiltonian-based ans\"atze as it is done above would lead to very long circuits.
Therefore, we do not consider VHA for this problem. 

In consequence, as a simple implementation of QOCA to a quantum chemistry problem, we use a variation of the ansatz based on the 12-qubit Hamiltonian of an open~$1\times 6$ Fermi-Hubbard chain with the drive terms of~\cref{eq: driveX,eq: driveY}. 
Although the water molecule Hamiltonian describes a richer set of fermionic interactions than the FHM, this choice of ansatz offers one of the simplest construction that simulates electron-electron correlations and is therefore a good starting point. 
Moreover, the ansatz is fully parametrized as before and the simulations were achieved under the same numerical conditions. 

The maximum fidelities achieved for the QOCA, HEA and UCCSD variational forms are reported in~\cref{tab: max fids H2O} for different number of ansatz layers~$d$ and initial states, which are either the Hartree-Fock (HF) approximation to the ground state or the equal superposition of all basis states~$\ket{+}^{\otimes N}$. 
Both initial states require one layer of single-qubit gates to prepare. 
The number of variational parameters $(n_\theta/d)$ and CNOT gates $(n_{\text{CX}}/d)$ per layer are also presented.
For the case of UCCSD we use~$d=1$, as it is proven to be enough for the simulation of chemical systems~\cite{romero2018strategies,o2016scalable}.
For this reason, UCCSD uses significantly fewer parameters than other approaches, however, the resulting circuit requires roughly the same two-qubit-gate count as a~$d=5$ QOCA circuit.

We observe that a single layer ($d=1$) QOCA circuit can prepare the ground state of the water molecule with fidelity~$0.9742$, a performance which is comparable to that of the well-established UCCSD ansatz ($0.9748$). 
However, while this does not improve the performance of UCCSD, adding layers up to $d=5$ for QOCA increased the fidelity to~$0.9931$. 
Interestingly, the~$\ket{+}^{\otimes N}$ initial state, which has a~$9.95\times 10^{-5}$ overlap with the target state, yields better results for QOCA with fewer ansatz layers than the Hartree-Fock initial state, which has a~$0.9735$ overlap. These simulations suggest that QOCA can be useful also for quantum-chemistry problems. Modifying the QOCA circuit to better reproduce the interactions between the spin orbitals of the water molecule could lead to further improvements in performance. 

\section{Conclusion}
\label{sec: conclusion}
We introduced the Quantum-Optimal-Control-ins\-pired Ansatz by adding carefully chosen symmetry-breaking drive terms to the problem Hamiltonian and parametrizing the resulting time-evolution-like operator.
We first applied QOCA to the half-filled Fermi-Hubbard model and found that in most cases it yields to a faster and more accurate convergence than standard approaches, even with unstructured initial states having little overlap with the target ground state.
We showed evidence that this improved convergence is made possible by the symmetry-breaking terms allowing for small excursions outside of the target symmetry sector of the problem Hamiltonian. 
Moreover, we used QOCA to prepare the ground state of the water molecule, and showed that it can surpass the commonly used UCCSD ansatz with drastically shorter circuits.

Its broader applicability and the flexibility in choosing drive terms make QOCA a promising approach to tackle a wide range of quantum chemistry and materials problems on near-term quantum computers.
Although the QOCA circuits are currently too deep to be implemented reliably on today's quantum devices, we expect that it may exhibit some resilience to symmetry-breaking errors.
Our work represents a first step towards the development of a more general class of symmetry-breaking ans\"atze for variational quantum algorithms.

\begin{acknowledgments}
    We thank David Poulin, Jonathan Gross and Alexandre Daoust for useful discussions. This work was undertaken thanks in part to funding from NSERC, the Canada First Research Excellence Fund and the U.S. Army Research Office Grant No. W911NF-18-1-0411.
\end{acknowledgments}

\paragraph*{Note added} After completion of this work, we became aware of related work that was recently posted~\cite{vogt2020preparing}.
 
\bibliography{biblio}

\appendix

\section{Jordan-Wigner fermionic encoding}
\label{sec: JW}

In the Jordan-Wigner transformation, each fermionic site is encoded into the state of two qubits with the mapping~$(0,\uparrow,\downarrow,\uparrow\downarrow)\mapsto(00,01,10,11)$. Moreover, the fermionic ladder operators take the form
\begin{align}\label{eq: JW}
    \ha_p &\mapsto \hat\sigma_p\, \bigotimes_{l<p}\,\hZ_l , \nonumber\\
    \had_p &\mapsto \hat\sigma^{\dagger}_p\, \bigotimes_{l<p}\,\hZ_l ,
\end{align}
where~$\hat\sigma = \ket{0}\bra{1}$,~$\hZ$ is the Pauli-$Z$ operator and the indices denote the spin orbitals or qubits.
For a lattice of~$L$ sites, we arrange the~$N=2L$ spin orbitals as~$\ket{f_{1\uparrow}\dots f_{L\uparrow} ; f_{1\downarrow}\dots f_{L\downarrow}}$ with~$f_p\in \{0,1\}$ the occupation of spin-orbital~$p$.

With this mapping, hopping terms between spin-orbitals~$p$ and~$q$ with \hbox{$p<q$} transform as
\begin{equation}
    \had_{p}\ha_{q} + \had_{q}\ha_{p} \mapsto \frac{1}{2} (\hX_p\hX_q + \hY_p\hY_q)\bigotimes_{l=p+1}^{q-1} \hZ_l,
\end{equation}
where~$\hX,\,\hY$ and $\hZ$ are Pauli matrices. 
The product of~$\hZ$ operators, referred to as the JW string, vanishes when~$q=p+1$. 
Moreover, the number operator on spin-orbital~$p$, and therefore the onsite Coulomb interaction between spin-orbitals~$p$ and~$q$ take the form
\begin{align}
    \hn_p = \had_p\ha_p &\mapsto \frac{1}{2} (\hat I-\hZ_p),  \nonumber\\
    \hn_p\hn_q &\mapsto \frac{1}{4} (\hat I-\hZ_p-\hZ_q+\hZ_p\hZ_q).
\end{align}
At half-filling, the single~$\hZ$s coming from the onsite interaction terms are canceled by similar terms arising from the chemical potential, leading to a simple expression for the potential 
\begin{equation}
    \hV \mapsto \frac{U}{4}\sum_{i=1}^{L} \hZ_{i\uparrow}\hZ_{i\downarrow},
\end{equation}
which is diagonal in the computational basis.

\section{Parametrization of the ans\"atze}
\label{sec: parametrization}

\subsection{Full parametrization} 
\label{sec: full param}
This strategy corresponds to taking all (or almost all) gate angles as variational parameters. 
This gives the classical optimizer enough freedom to explore the Hilbert space spanned by the ansatz at the cost of a longer optimization time.  
We note that the HEA has, by default, a \textit{fully parametrized} configuration since all single-qubit gates are parametrized. 
Moreover, the same strategy for VHA consists of assigning one parameter to every~$\had_{i\sigma}\ha_{j\sigma} + \text{h.c.}$ hopping terms and duplicating the parameter to take into account the two spin orientations. 
This is because at half-filling and zero total spin, there is a spin-inversion symmetry which removes the need to treat spins up and down differently. 
Additionally, every term of the on-site interaction is associated with a variational parameter. 
The asymptotic scaling of number of variational parameters for all ans\"atze is summarized in~\Cref{tab: parameters} for both parametrization strategies.

\subsection{Scalable parametrization} 

In a scalable parametrization strategy, we employ a number of variational parameters that is independent of the system size. 
Because there are fewer parameters, we expect the optimization to be faster, but larger circuit depths might be necessary to achieve the same accuracy as full parametrization. 

Although it is less clear how one would achieve a scalable parametrization for hardware-efficient approaches, a simple strategy exists for physics-inspired ans\"atze such as QOCA. 
It consists in grouping the individual terms of the Hamiltonian into a constant number of sets containing commuting terms.
For example, a common way of grouping the different terms of the FHM on a 2D lattice is
\begin{equation}
    \hH_{\text{FHM}} = \hH_{h,\text{even}} + \hH_{h,\text{odd}} + \hH_{v,\text{even}} + \hH_{v,\text{odd}} + \hH_U,
\end{equation}
where the first four terms now group the even and odd, vertical and horizontal hopping terms, while~$\hH_U$ collects the on-site interaction terms. 
Note that for the 3D FHM, two additional sets of hopping terms covering the third dimension would be necessary.

\begin{table}[h]
    \centering
    \begin{tabular}{@{}lcc@{}}
     & \begin{tabular}[c]{@{}c@{}}Full \\ parametrization\end{tabular} & \begin{tabular}[c]{@{}c@{}}Scalable \\ parametrization\end{tabular} \\ \midrule\midrule
    \multicolumn{1}{l}{HEA} &~$2Ld$ & -- \\
    \multicolumn{1}{l}{VHA} &~$(\eta+1)Ld$ &~$(2\eta+1)d$ \\
    \multicolumn{1}{l}{FT-VHA} &~$(\eta+1)Ld$ &~$(\eta+1)d$ \\
    \multicolumn{1}{l}{QOCA} &~$(\eta+3)Ld$ &~$(2\eta+3)d$ \\
    \multicolumn{1}{l}{sQOCA} &~$3Ld$ &~$3d$ \\ \bottomrule
    \end{tabular}
    \caption{\label{tab: parameters} Asymptotic scaling of the number of variational parameters of the ans\"atze of this work for the full and scalable parametrization strategies. 
    These numbers are for periodic~$\eta$-dimensional Fermi-Hubbard systems of~$L$ lattice sites.~$d$ is the number of layers of the ans\"atze.}
\end{table}

\section{Numerical simulations}
\label{sec: numerical}

All simulations are done using Qiskit Aqua's VQA tools~\cite{Qiskit}. 
Because noise is not considered, a unitary statevector simulator is used. For simplicity, we also assumed all-to-all connectivity of the qubits, although this is not strictly needed. 
We chose the COBYLA~\cite{powell1994direct,powell1998direct,powell2007view} method as the classical optimizer with a maximum number of function evaluation of~$\sim 10^5$. 
This number was justified as being reasonable in~\cite{cade2019strategies} using experimentally realistic arguments. 

Whenever possible, we initialize all variational para\-meters to zero. 
With this choice, Hamiltonian-based ans\"atze implement the identity operator at the start of the optimization routine and the variational search begins from the initial state. 
In contrast to a random initialization of the parameters, this strategy also avoids the need of doing repeated VQA runs and post-selecting the best results.
However, in the case of short-QOCA, this strategy results in premature convergence of the optimizer into states close to the initial guess, forcing us to use a random initialization of the parameters. 
Interestingly, even without post-selection, this did not hinder the convergence capability thanks to the robustness of QOCA regarding initial conditions.

Finally, all layers of the ans\"atze are optimized simultaneously. 
Further improvement can potentially be achieved by adopting a layer-by-layer optimization strategy as in Ref.~\cite{wecker2015progress}.

For the simulation of the water molecule, we use the PySCF driver to obtain the Hamiltonian as provided 

\section{Initial states}
\label{app: initial states}

In most quantum simulations of the FHM reported in the literature~\cite{wecker2015progress,dallaire2019low,reiner2019finding,verdon2019learning,xu2020test,cade2019strategies,montanaro2020compressed}, the initial state is the ground state of the non-interacting FHM \ie fixing~$U=\mu=0$ in~\cref{eq: FHM}. 
Because the resulting Hamiltonian is diagonal in Fourier space, this is a convenient choice because the ground state is readily computed classically. 
However, preparing this on a quantum computer generally requires very long quantum circuits as it involves the fermionic Fourier transformation.
Current implementations of this transformation~\cite{verstraete2009quantum,jiang2018quantum,babbush2018low} are defined only for periodic systems, which limits this initial state's applicability.
To the best of our knowledge, no implementation of an open-boundary-conditions fermionic Fourier transformation has been developed to date.
Furthermore, the ground state of the non-interacting FHM can be degenerate which makes it difficult to choose which one or superposition thereof to use.
This challenge is often pointed out as an open problem~\cite{cade2019strategies,xu2020test}, since in most VQA realization, prior knowledge of the target state is used to find the initial state that maximizes the fidelity.
It is unclear how one would make this choice as systems grow computationally intractable.

\subsection{The non-interacting Fermi-Hubbard model}
To see how this degeneracy arises, we consider the 1D non-interacting FHM ($U=\mu=0$) with~$L$ sites and periodic boundary conditions. 
In momentum space, the Hamiltonian is given by a collection of free fermionic modes
\begin{equation}\label{eq: Tk}
    \hat{\mathcal{T}} = \text{FT}\ \hT\ \text{FT}^\dagger = \sum_{k,\sigma=\{\uparrow,\downarrow \}} \varepsilon_k \hcd_{k\sigma} \hc_{k\sigma},
\end{equation} 
where the energy spectrum is 
\begin{equation}
    \label{eq: fourier energy}
    \varepsilon_k = -2t \cos \left(\frac{2\pi k}{L} \right).
\end{equation}
In the above Hamiltonian,~$\hcd_{k\sigma}$ and~$\hc_{k\sigma}$ are respectively the creation and annihilation fermionic operators of momentum~$k$ and spin~$\sigma$. 
They are obtained from the real-space ladder operators~$\had_{k\sigma}$ and~$\ha_{k\sigma}$ and the fermionic Fourier transformation as 
\begin{align}
    \hcd_{k\sigma} &= \text{FT}\ \had_{k\sigma}\ \text{FT}^\dagger = \frac{1}{\sqrt{L}}\sum_{j=0}^{L-1} e^{-i\frac{2\pi k}{L}j} \had_{j\sigma},\\
    \hc_{k\sigma} &= \text{FT}\ \ha_{k\sigma}\ \text{FT}^\dagger = \frac{1}{\sqrt{L}}\sum_{j=0}^{L-1} e^{i\frac{2\pi k}{L}j} \ha_{j\sigma}.
\end{align}

Because~$k$ can only take discrete values, one notices that a degeneracy appears when there are energy levels at~$\varepsilon_k=0$ since these levels could be occupied or empty without affecting the ground state energy. 
It is straightforward to see from~\cref{eq: fourier energy} that this can happen only when~$L= 4l$, with~$l$ an integer. 
In this case, there are two values of~$k$ (corresponding to~$k=l$ and~$k=3l$) which leads to~$\varepsilon_k=0$. 
The degeneracy is therefore \hbox{$4^{2} = 16$} since each momentum mode can be empty, occupied by a~$\uparrow$ or~$\downarrow$ spin, or both. 
In the half-filled symmetry sector, the degeneracy is reduced to~$\left(\begin{smallmatrix}4 \\ 2\end{smallmatrix}\right) = 6$. 
Note that in the case~$L\neq 4l$, the ground state of the non-interacting FHM is not degenerate and is a simple basis state in momentum space.

As mentioned above, this occasional degeneracy makes it difficult to guess which basis state (or superposition thereof) is the best initial state to use in a VQA. 
Although, one can select states that respect certain desired properties such as particle number, total spin and total momentum. 

Typically, the degeneracy at~$L= 4l$ can be lifted by applying a small perturbative Coulomb interaction $U$. 
In this case, the ground state of the non-interacting FHM becomes a superposition of basis states in Fourier space. 
One must apply the FT$^\dagger$ in order to transform this initial
state into real space for the VQA.

\subsection{Choosing and preparing the initial states}
In the case of~$L= 4$ (or $2\times 2$), we computed the fidelity of the 16 degenerate ground states of~\cref{eq: Tk} with respect to the target ground state and post-selected the ones leading to the highest fidelity.
This strategy is, of course, not scalable and therefore it remains unclear how one would proceed in practice in the case where the fidelity with the target ground state cannot be computed beforehand.

In the present case, this strategy yields two ground states with a fidelity of~$\approx 0.425$ with respect to the ground state of the full model. 
Labeling the spin orbitals~$\ket{f_{1\uparrow}\dots f_{L\uparrow} ; f_{1\downarrow}\dots f_{L\downarrow}}$, these two states in real space are 
\begin{align}
    \ket{\Omega_T^{(1)}} &= \text{FT}^\dagger\ \ket{1100\ ;1100},\label{eq: 1GS_1}\\
    \ket{\Omega_T^{(2)}} &= \text{FT}^\dagger\ \ket{1001\ ;1001}.\label{eq: 1GS_2}
\end{align}
Preparing these two states requires applying Pauli-$X$ gates on selected qubits followed by the fermionic Fourier transformation, something which requires long quantum circuits~\cite{verstraete2009quantum,jiang2018quantum,babbush2018low}.

Adding a small perturbation~$U=1\times 10^{-5} t$, we find that the following superposition of~$\ket{\Omega_T^{(1)}}$ and~$\ket{\Omega_T^{(2)}}$ yields a significantly larger fidelity to the true ground state of~$\approx 0.85$:
\begin{align}\label{eq: GST}
    \ket{\Omega_T} &= \frac{\ket{\Omega_T^{(1)}}-\ket{\Omega_T^{(2)}}}{\sqrt{2}}\nonumber \\
     &= \text{FT}^\dagger\ \frac{\ket{1100\ ;1100}-\ket{1001\ ;1001}}{\sqrt{2}}.
\end{align}
This, however, increases the complexity of the initial state preparation.

\end{document}